\documentclass[11pt]{article}
\usepackage{amssymb,amsmath}
\usepackage{slashed}
\usepackage{mathtools}
\usepackage{feynmf}

\usepackage[T1]{fontenc}
\usepackage[utf8]{inputenc}
\usepackage[english]{babel}

\usepackage{mathrsfs}
\usepackage{graphicx}
\usepackage{simplewick}
\usepackage{simpler-wick}
\usepackage{amsxtra}

\usepackage{footnote}
\usepackage{tablefootnote}
\usetikzlibrary{graphs,quotes,arrows.meta}

\usepackage[export]{adjustbox}
\usepackage[colorinlistoftodos]{todonotes}
\usepackage[colorlinks=true, allcolors=blue]{hyperref}
\usepackage{hyperref}
\usepackage{authblk}

\usepackage{tikz}
\usetikzlibrary{graphs}
\usepackage{subfigure}
\usepackage{indentfirst}
\usepackage{textcomp}
\usepackage{gensymb}
\usetikzlibrary{shapes.geometric, arrows}
\usepackage{csquotes}

\title{Consideration of a loop decay of dark matter particle into electron-positron from point of view of possible FSR suppression}

\author{Belotsky K.M.\\k-belotsky@yandex.ru
\\Kamaletdinov A.Kh.\\kamaletdinov.a.h@yandex.ru}
\affil{National Research Nuclear University MEPhI (Moscow Engineering Physics Institute), 115409, Kashirskoe shosse 31, Moscow, Russia}
\begin{document}
\maketitle

\begin{abstract}
Cosmic positron anomaly is still not explained. Explanation with dark matter (DM) decay or annihilation is one of the main attempts to do it. But they suffer with shortcoming as overproduction of induced gamma-radiation which contradict to cosmic gamma-background. Final state radiation (FSR) in such processes is supposed under standard conditions (by default) to have the basic contribution in it. Our group elaborates possibility to evade this problem in different ways. Here we continue one of them connected with possibility of suppression of FSR due to specifics of Lagrangian describing DM particle decay. Loop through two new spinors and scalar is considered. Effect of FSR suppression is found to be existing but at the very low level in the considered case.
\end{abstract}



\section{Introduction\label{s:intro}}

The problem of Dark Matter (DM) is one of the main long-term problems of fundamental physics. Many direct and indirect searches for DM particles are undertaken. Cosmic rays (CR) relate to the indirect one and the revealed cosmic positron anomaly (PA) \cite{ref:Pamela, ref:AMS1, ref:AMS2, ref:Ackerman, ref:FermiLat0}  can be supposed to be a possible indication of DM. 

But attempts to explain the positron anomaly with DM face a problem of agreement with data on cosmic gamma-rays (see, e.g., our works \cite{ref:FermiLatKills, ref:GRExplanation, ref:ExcessVSIGRB} and other \cite{ref:Laletin, ref:FermiOther1, ref:FermiOther2, ref:FermiOther3, ref:FermiOther4}) and CMB \cite{ref:PlanckResults} and some other for specific DM model case. Constraint following from CMB can be more easily avoided (see, e.g., the references in \cite{ref:IndirectEffectsOfDM, ref:CosmicGammaRayConstraints, ref:DampeExcess}) than that from data on cosmic gamma-ray background \cite{ref:FermiLat}. This constraint seems to be the least model dependent. When high energy positrons and electrons $e^{\pm}$ are produced from DM decay or annihilation, it will be accompanied by final state radiation (FSR) and they will scatter on medium photons. Both processes give us gamma of high energy.

The most popular alternative approach to the solution of the problem of PA origin is associated with nearby pulsars. But it also strongly constrained (if not excluded) by data on gamma-radiation \cite{ref:GRPulsars1, ref:GeVObservations, ref:GRPositrons}.  So the question of PA origin is still open. 

We consider possibility of PA explanation with the help of DM and elaborate two approaches for it: one is connected with space distribution of DM in Galaxy ('Dark disk' model) \cite{ref:FermiLatKills, ref:ExcessVSIGRB, ref:GRExplanation, ref:Laletin, ref:DarkDisk1, ref:DarkDisk2}, other one is connected with possible physics of DM interaction leading to annihilation or decay which can give suppressed FSR \cite{ref:IndirectEffectsOfDM, ref:CosmicGammaRayConstraints, ref:DampeExcess}. The latter was attempted to be considered by other recently \cite{ref:Gravitino}. Here we make one more step in this investigation. We study one more decay mode of DM particle which contains a loop from spinor and scalar particle of dark sector. The process is drawn below. The obtained answer is that the effect is negligible in the considered case, though it exists in principle, i.e. relative probability of FSR photon production can be changed.

Below we present theoretical initial settings for interaction/decay physics of DM particles and basic calculations, then conclude.


\section{DM decay model considered and calculation details}


Let us consider two processes of DM particle ($X$) decay:
$X \rightarrow e^+ e^-$ and the same with FSR $X \rightarrow e^+ e^- \gamma$.
The goal of the task is to minimize the ratio:
\begin{equation}
    {\Gamma(X \rightarrow e^+ e^- \gamma) \over \Gamma(X \rightarrow e^+ e^-)} = \mathrm{min},
    \label{eq:ratio}
\end{equation}
 where $\Gamma(X \rightarrow e^+ e^- (\gamma))$ are the respective decay widths.
 
In order to be able to see the photon suppression at different energies, we study the energy distribution of the photon emission probability in the decays of DM particles (\ref{eq:distr}).

\begin{equation}
    \frac{\partial Br(e^+ e^- \gamma)}{\partial \omega} \equiv \frac{\partial}{\partial \omega} \bigg( \frac{\Gamma(X \rightarrow e^+ e^- \gamma)}{\Gamma(X \rightarrow e^+ e^-)} \bigg),
    \label{eq:distr}
\end{equation}
where $\omega$ is the energy of the final state photon.
 
As was shown earlier \cite{ref:IndirectEffectsOfDM}, the simplest interaction vertices such as (\ref{eq:simplestVertices1}, \ref{eq:simplestVertices2}) do not lead to a significant suppression of the photon yield in a such decays. These are
\begin{equation}
    \mathscr{L}_{scalar} = X\overline{\psi}(a+b\gamma^5)\psi, \newline \quad \mathscr{L}_{vector} = \overline{\psi}\gamma^\mu (a+b\gamma^5)X_\mu \psi,
\label{eq:simplestVertices1}
\end{equation}
\begin{equation}
    \mathscr{L}_{C} = X\overline{\psi^C} (a+b\gamma^5)\psi, \qquad \mathscr{L} = \overline{\psi}\gamma^{\mu}(a+\frac{b(\gamma^{\nu}\partial_{\nu})}{m})X_{\mu}\psi.
    \label{eq:simplestVertices2}
\end{equation}
Also shown that complication of process kinematics does not give an effect \cite{ref:DampeExcess, ref:FSRSuppressionMass}.

\begin{figure}[h!]
    \centering
    \includegraphics[scale=0.55]{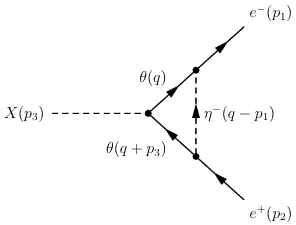}
    \caption{Feynman diagram of two body decay process.}
    \label{fig:twobody}
\end{figure}

Here we consider one of the options for complicating the DM-SM interactions. On the base of previous works we suppose that it is worth to consider other type of the processes. Loop diagrams of DM decays into $e^+, e^-$ particles can be worth to be studied. We consider here the interaction Lagrangian of the form (\ref{eq:Lagr}):

\begin{equation}
    \mathcal{L}_{\triangle} = X \Bar{\theta}(a+\mathrm{i}\:b\gamma^5)\theta + \eta\; \Bar{\theta} (c+\mathrm{i}\:d\gamma^5) \Psi + \eta^*\; \Bar{\Psi} (c+\mathrm{i}\:d\gamma^5) \theta,
    \label{eq:Lagr}
\end{equation}
where $\theta$ is considered as the fermionic neutral DM component, and $X, \eta$ -- as the scalar DM particles.
In this work, to simplify the calculations, the mass of the $\eta$ particles is assumed to be very large so that the photon emission by the $\eta$ propagator can be neglected.
The leading order of the  process $X \rightarrow e^+ e^-$ in such case describes by triangle-loop diagram shown in figure \ref{fig:twobody}.

We evaluate the corresponding matrix element (\ref{eq:ME}) here through the form-factors $F_1$ and $F_2$ using the Passarino-Veltman (PV) reduction procedure, described in \cite{ref:PV1, ref:PV2}. In order to perform calculations with PV-functions the PackageX \cite{ref:PackageX} tool for Wolfram Mathemetica was used. Matrix element is

\begin{equation}
    i \; \mathcal{M} = i \; \Bar{u}(p_1) \Big( F_1(\sqrt{s}) - i F_2(\sqrt{s}) \gamma^5 \Big) v(p_2),
    \label{eq:ME}
\end{equation}

\begin{equation}
\begin{gathered}
    F_1(\sqrt{s}) = a(c^2 - d^2) \Big( B(\sqrt{s}) + (m_2^2 + m_1 m_3) C_0(p_1,p_2) \Big)
    + \\ +
    2 b c d \Big( B(\sqrt{s}) + (m_2^2 - m_1 m_3) C_0(p_1,p_2) \Big),
    \\
    F_2(\sqrt{s}) = b(c^2 - d^2) \Big( B(\sqrt{s}) + (m_2^2 - m_1 m_3) C_0(p_1,p_2) \Big)
    - \\ -
    2 a c d \Big( B(\sqrt{s}) + (m_2^2 + m_1 m_3) C_0(p_1,p_2) \Big).
\end{gathered}
\end{equation}

We use here and further notation $B(\sqrt{s}) \equiv B_0(\sqrt{s};\; m_1, m_3)$, $C_i(p_1, p_2) \equiv C_i(p_1, p_2;\; m_1, m_2, m_3)$.
In this case, the squared amplitude of the two-body decay averaged over the final state polarizations takes the form:
\begin{equation}
    \frac{1}{4}\sum_{\lambda} \mathcal{M} \mathcal{M^*} = \frac{m_X^2}{2}\Big( F_1(\sqrt{s})^2 + F_2(\sqrt{s})^2 \Big),
\end{equation}

After the same calculations carried out for the three-body decay process $(X \rightarrow e^+ e^- \gamma)$ (see figure \ref{fig:threebody}) and looking at their ratio one can obtain the expression for final-state photon yield energy distribution (\ref{eq:distr}):
\begin{figure}[h!]
    \centering
    \includegraphics[scale=0.3]{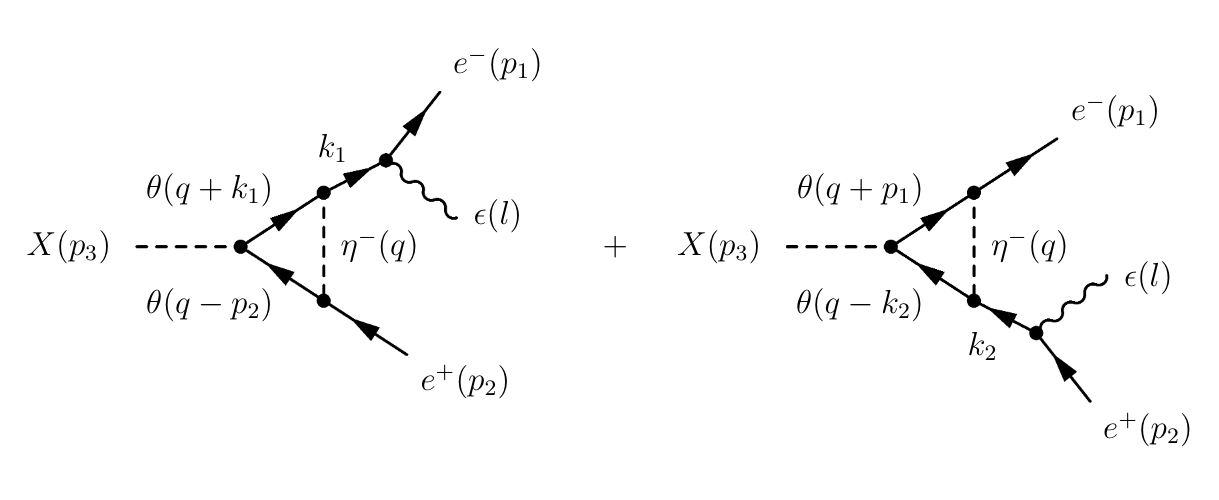}
    \caption{Feynman diagrams of three body decay process.}
    \label{fig:threebody}
\end{figure}

\begin{equation}
    \frac{1}{4}\sum_{\lambda} \mathcal{M} \mathcal{M^*} = \frac{(c^2 + d^2)^2}{4} \Big(A_{11} - A_{12} - A_{21} + A_{22} \Big),
\end{equation}

\begin{equation}
\begin{gathered}
    A_{ii} = a^2 \frac{|X_i^+|^2 + 2 m_1^2 (l \cdot p_i)^2 (p_1 \cdot p_2) |Y_i|^2}{(p_i+l)^4}
    + \\ +
    b^2\frac{|X_i^-|^2 + 2 m_1^2 (l \cdot p_i)^2 (p_1 \cdot p_2) |K_i|^2}{(p_i+l)^4},\\
\end{gathered}
\end{equation}

\begin{equation}
\begin{gathered}
    A_{ij} = 2 m_1^2 (p_1\cdot p_2) (l\cdot p_1)(l\cdot p_2) \frac{a^2 Y_i Y_j^* - 4 a^2 C_i C_j^* + b^2 K_i K_j^*}{(p_i + l)^2 (p_j + l)^2} - \\ - \Big(p_1 \cdot (p_2 + l)\Big)\Big(p_2 \cdot (p_1 + l)\Big) \frac{a^2 X_i^+ X_j^{+*} + b^2 X_i^- X_j^{-*}}{(p_i + l)^2 (p_j + l)^2}
\end{gathered}
\end{equation}
where $K_1 = C_0(k_1, p_2), \; K_2 = C_0(p_1, k_2), \; C_1 = C_1(k_1, p_2), \; C_2 = C_1(p_1, k_2)$,

\begin{equation}
\begin{gathered}
    X_1^{\pm} = 2 (l \cdot p_1) C_1 + B(\sqrt{s}) + K_1 (m_2^2 \pm m_1^2),
    \\
    X_2^{\pm} = 2 (l \cdot p_2) C_2 + B(\sqrt{s}) + K_2 (m_2^2 \pm m_1^2),
    \\
    Y_1 = 2 C_1 + K_1 \qquad Y_2 = 2 C_2 + K_2.
\end{gathered}
\end{equation}
 
 The study of the influence of model parameters variation on the photon emission showed that the suppression turns out to be insignificant in order to explain satisfactorily the high energy cosmic positron spectrum not contradicting to gamma-ray background.

\begin{figure}[h!]
    \centering
    \includegraphics[scale=0.35]{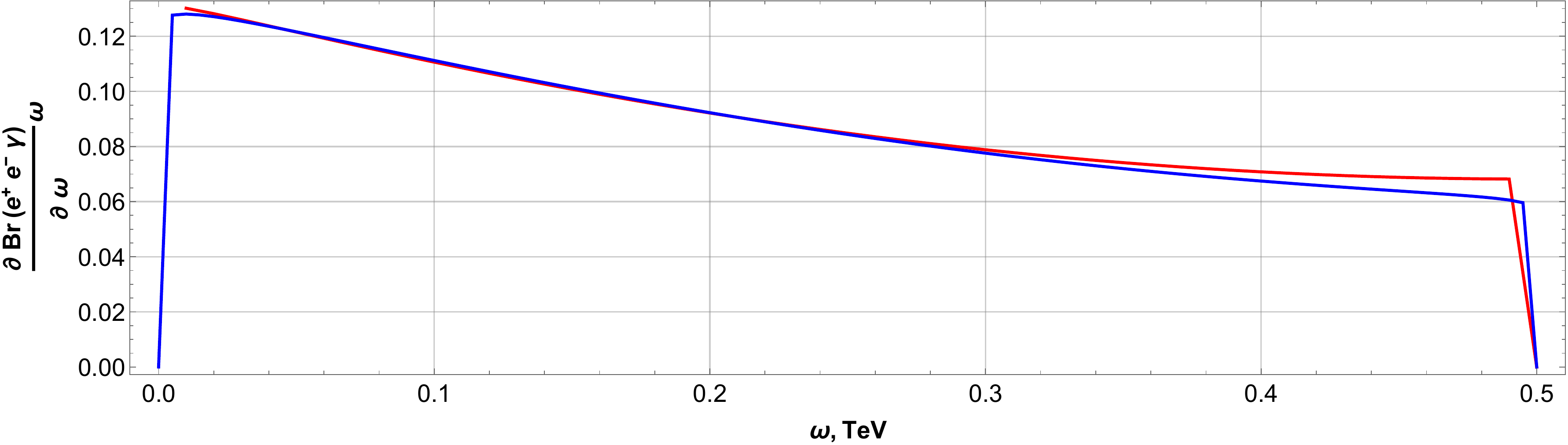}
    \caption{The comparison of the photon yields in the case of simplest scalar vertice (\ref{eq:simplestVertices1}) (blue) and the loop vertice (red)}
\end{figure}

\section{Conclusion}
In this paper we continue our studying possibility to suppress FSR (gamma emission) in DM explanation of cosmic positron anomaly. Here we consider specific DM-lepton interaction Lagrangian which allows decaying DM particle to $e^{\pm}$ through the loop of intermediate particles of dark sector. We obtained relative probability of FSR production (branching ratio of the respective mode) in dependence of final photon energy analytically up to the level of the squared matrix element. It is important for understanding whether or not possible FSR suppression looking at dependence on model parameters at high photon energies (most critical) and prospectiveness of possible complication of the model. Now it is obtained that the considered variant of the loop decay is not able to facilitate solution PA origin with DM, though shows (maybe, opens new) principal opportunity of FSR yield changing.

\section*{Acknowledgements}
The work was supported by the Ministry of Science and Higher Education of the Russian Federation
by project No 0723-2020-0040 ``Fundamental problems of cosmic rays and dark matter''.
Also, we would like to thank M.Solovyov for the help in providing by references.

\newpage

\end{document}